\documentclass[letterpaper]{article}

\newcommand{\D}{\Delta}
\newcommand{\dd}{\mathrm{d}}
\newcommand{\mo}{{\cal O}}
\newcommand{\mA}{\mathcal{A}}
\newcommand{\an}[1]{\left\langle#1\right\rangle}
\newcommand{\sq}[1]{\left[#1\right]}
\newcommand{\zb}{\bar{z}}
\newcommand{\hb}{\bar{h}}
\newcommand{\lam}{\lambda}
\newcommand{\lamh}{\hat{\lambda}}
\newcommand{\lamt}{\tilde{\lambda}}

\usepackage{amsmath}
\usepackage{tikz}
\usepackage{jheppub}

\title{On Effective Field Theories with Celestial Duals}
\author[a]{Lecheng Ren,}
\author[a,b]{Marcus Spradlin,}
\author[a]{Akshay Yelleshpur Srikant}
\author[a]{and Anastasia Volovich}

\affiliation[a]{Department of Physics,
	Brown University,
	Providence, RI 02912, USA}
\affiliation[a]{Brown Theoretical Physics Center,
	Brown University,
	Providence, RI 02912, USA}

\emailAdd{lecheng\_ren@brown.edu}
\emailAdd{marcus\_spradlin@brown.edu}
\emailAdd{akshay\_yelleshpur\_srikant@brown.edu}
\emailAdd{anastasia\_volovich@brown.edu}

\abstract{We show that associativity of the tree-level OPE in a celestial CFT imposes constraints on the coupling constants of the corresponding bulk theory. These constraints are the same as those derived in arXiv:2111.11356 from the Jacobi identity of the algebra of soft modes. The constrained theories are interesting as apparently well-defined celestial CFTs with a deformed $w_{1+\infty}$ symmetry algebra. We explicitly work out the ramifications of these constraints on scattering amplitudes involving gluons, gravitons and scalars in these theories. We find that all four-point amplitudes constructible solely from holomorphic or anti-holomorphic three-point amplitudes vanish on the support of these constraints, which implies that all purely holomorphic or purely anti-holomorphic higher-point amplitudes vanish.}

\begin{document}

\maketitle

\section{Introduction}

The central tenet of celestial holography~\cite{Pasterski:2016qvg,Pasterski:2017kqt,Pasterski:2017ylz} is that a scattering amplitude of $n$ massless particles, with momenta $\{p_i\}$ and helicities $\{s_i\}$, when recast in a basis of boost eigenstates, can be interpreted as a correlation function of $n$ operators with conformal weights $(h_i, \hb_i)$ in a two-dimensional celestial conformal field theory (CCFT). The four-dimensional helicity $s_i$ of particle $i$ is the same as the two-dimensional spin $s_i=h_i-\hb_i$, and we use $\D_i = h_i+\bar{h}_i$.

This dictionary implies that the OPE of two operators in CCFT is determined by the collinear limit of the corresponding scattering amplitudes~\cite{Fan:2019emx, Pate:2019lpp, Himwich:2021dau}. Specifically, the OPE is given by
\begin{align}
    \label{eq:genericOPE}
    \mo_{h_1, \bar{h}_1}(z_1, \zb_1) \mo_{h_2, \bar{h}_2} (z_2, \zb_2)\sim \frac{1}{z_{12}} \sum_{p} \sum_{m=0}^{\infty} C_p^{(m)} \left(\bar{h}_1, \bar{h}_2\right)\zb_{12}^{p+m}\,\bar{\partial}^{m} \mo_{h_1+h_2-1, \bar{h}_1+\bar{h}_2+p} (z_2, \zb_2)
\end{align}
where $z_{ij}=z_i-z_j$ and the OPE coefficient $C_p^{(m)} \left(\bar{h}_1, \bar{h}_2\right)$ corresponding to the contribution of an operator with weights $(h_1+h_2-1, \bar{h}_1+\bar{h}_2+p+m)$ is given by~\cite{Fan:2019emx,Pate:2019lpp,Himwich:2021dau}
\begin{align}
    \label{eq:genericopecoefficient}
     C_p^{(m)} (\bar{h}_1, \bar{h}_2)  &=-\frac{1}{2} \kappa_{s_1, s_2,-s_I} \frac{1}{m!} B(2\bar{h}_1 + p + m, 2\bar{h}_2 + p),
\end{align}
where $B(a,b)$ is the Euler beta function and $\kappa_{s_1,s_2,-s_I}$ is the coupling constant appearing in the bulk three-point scattering amplitude of particles with helicities $s_1$, $s_2$ and $-s_I$, with $s_I = s_1+s_2-p-1$. Thus there is a direct link between the bulk three-point couplings and the celestial OPE. Restricting to tree-level and excluding massless higher spins leaves a finite roster of bulk three-point amplitudes which could potentially contribute to the sum in~(\ref{eq:genericOPE}). These have been tabulated in~\cite{Pate:2019lpp}.

It was observed in~\cite{Guevara:2021abz} that the OPE of conformally soft graviton operators (those with $\D_i = 2, 1, 0, -1, -2, \dots$), in the absence of higher derivative interactions in the bulk, forms a symmetry algebra. This algebra was identified as {the Kac Moody algebra of the wedge subalgebra of $w_{1+\infty}$} in~\cite{Strominger:2021lvk}. The inclusion of higher derivative interactions in the bulk deforms the algebra as demonstrated in~\cite{Mago:2021wje}. { In particular, note that for a given value of $p$, only the soft currents with $\D \geq p-1$ are modified. This implies that there is short list of operators that can modify the subleading and subsubleading soft currents. This is consistent with the results of~\cite{He:2014bga,DiVecchia:2016amo,Elvang:2016qvq}}. In~\cite{Mago:2021wje} it was also shown that the algebra of soft modes violates the Jacobi identity unless the bulk couplings satisfy certain particular constraints (reviewed in Section~\ref{sec:softOPEassoc} below). In particular, this implies that the OPE of the soft modes is not associative for arbitrary couplings. It is natural to wonder about the implications of celestial OPE associativity on bulk scattering amplitudes. The purpose of this paper is to answer this question and to investigate the properties of amplitudes in theories which have an associative celestial OPE at tree level.

The paper is organized as follows. In Section~\ref{sec:softOPEassoc} we briefly review the constraints obtained by imposing the Jacobi identity on the algebra of soft modes~\cite{Mago:2021wje}. We then rephrase the question of tree-level associativity directly at the level of bulk scattering amplitudes in Section~\ref{sec:hardOPEassoc} and show that they yield the same constraints on the bulk couplings. In Section~\ref{sec:implications} we demonstrate that these constraints lead to the vanishing of certain four-point amplitudes and we work out the consequences on amplitudes of higher multiplicity. We close with a discussion of various open questions.

\section{Review of the coupling constants constraints}
\label{sec:softOPEassoc}

We begin by listing all relevant three-point amplitudes and briefly reviewing the constraints among their couplings found in~\cite{Mago:2021wje}. The most generic scenario involves a graviton, gluons and scalars. However, there are subsets of these particles which yield an associative OPE.

The first subset is the graviton-scalar sector, where we consider the usual Einstein-Hilbert term with coupling proportional to $\kappa_{-2,2,2}$, an $R^3$ interaction proportional to $\kappa_{2,2,2}$, an $R^2 \phi$ interaction proportional to $\kappa_{0,2,2}$ and an $R \phi^2$ interaction proportional to $\kappa_{0,0,2}$. To be more precise, we specify this sector by the anti-holomorphic\footnote{We call $\mA(1^{s_1},2^{s_2},3^{s_3})$ anti-holomorphic if $s_1+s_2+s_3>0$ and holomorphic if $s_1+s_2+s_3<0$; see~(\ref{eq:threepoints}).} three-point amplitudes
\begin{equation}\label{eq:effgrav3}
    \begin{aligned}
        & \begin{tikzpicture}[baseline={([yshift=-0.9ex]current bounding box.center)},scale=0.8]
            \draw[thick] (-1.5,0) -- (-2.25,1.73*3/4) node[left]{$2^{++}$};
            \draw[thick] (-1.5,0) -- (0,0) node[above]{$3^{--}$};
            \draw[thick] (-1.5,0) -- (-2.25,-1.73*3/4) node[left]{$1^{++}$};
            \filldraw[thick,fill=white] (-1.5,0) circle (0.7) node{$\kappa_{-2,2,2}$};
        \end{tikzpicture} = \kappa_{-2,2,2} \frac{\lbrack 12 \rbrack^6}{\lbrack 23 \rbrack^2 \lbrack 13 \rbrack^2} \quad
        && \begin{tikzpicture}[baseline={([yshift=-0.9ex]current bounding box.center)},scale=0.8]
            \draw[thick] (-1.5,0) -- (-2.25,1.73*3/4) node[left]{$2^{++}$};
            \draw[thick] (-1.5,0) -- (0,0) node[above]{$3^{++}$};
            \draw[thick] (-1.5,0) -- (-2.25,-1.73*3/4) node[left]{$1^{++}$};
            \filldraw[thick,fill=white] (-1.5,0) circle (0.6) node{$\kappa_{2,2,2}$};
        \end{tikzpicture} = \kappa_{2,2,2} \lbrack 12 \rbrack^2 \lbrack 23 \rbrack^2 \lbrack 13 \rbrack^2 \\
        & \begin{tikzpicture}[baseline={([yshift=-0.9ex]current bounding box.center)},scale=0.8]
            \draw[thick] (-1.5,0) -- (-2.25,1.73*3/4) node[left]{$2^{++}$};
            \draw[thick] (-1.5,0) -- (0,0) node[above]{$3^\phi$};
            \draw[thick] (-1.5,0) -- (-2.25,-1.73*3/4) node[left]{$1^{++}$};
            \filldraw[thick,fill=white] (-1.5,0) circle (0.6) node{$\kappa_{0,2,2}$};
        \end{tikzpicture} = \kappa_{0,2,2} \lbrack 12 \rbrack^4 \quad
        && \begin{tikzpicture}[baseline={([yshift=-0.9ex]current bounding box.center)},scale=0.8]
            \draw[thick] (-1.5,0) -- (-2.25,1.73*3/4) node[left]{$2^\phi$};
            \draw[thick] (-1.5,0) -- (0,0) node[above]{$3^\phi$};
            \draw[thick] (-1.5,0) -- (-2.25,-1.73*3/4) node[left]{$1^{++}$};
            \filldraw[thick,fill=white] (-1.5,0) circle (0.6) node{$\kappa_{0,0,2}$};
        \end{tikzpicture} = \kappa_{0,0,2} \frac{\lbrack 12 \rbrack^2 \lbrack 13 \rbrack^2}{\lbrack 23 \rbrack^2}
    \end{aligned}
\end{equation}
together with their holomorphic parity conjugates. Note that in this paper we don't equate the coupling constants of the parity conjugate amplitudes with those of the original amplitude. Namely we regard $\kappa_{s_1,s_2,s_3} \neq \kappa_{-s_1,-s_2,-s_3}$ in general. The OPE of conformally soft operators satisfies the Jacobi identity only if the couplings are related by
\begin{align}
\label{eq:constraint11}
     (\kappa_{-2,2,2} - \kappa_{0,0,2})\, \kappa_{0,2,2} &= 0\,, \qquad (\kappa_{-2,2,2} - \kappa_{0,0,2})\, \kappa_{0,0,2} = 0 \,,\\
    &\label{eq:constraint12}  3\kappa_{0,2,2}^2 = 10\, \kappa_{-2,2,2}\, \kappa_{2,2,2}\,.
\end{align}
In particular, note that the presence of a scalar is required if the $R^3$ interaction has a non-zero coefficient.  The equivalence principle requires $\kappa_{-2,2,2}=\kappa_{0,0,2}$, which automatically ensures~(\ref{eq:constraint11}), but~(\ref{eq:constraint12}) is more nontrivial.

The second subset is the gluon-scalar sector defined by the three-point amplitudes
\begin{align}
    & \begin{tikzpicture}[baseline={([yshift=-0.9ex]current bounding box.center)},scale=0.8]
        \draw[thick] (-1.5,0) -- (-2.25,1.73*3/4) node[left]{$2^{b,+}$};
        \draw[thick] (-1.5,0) -- (0,0) node[above]{$3^{c,-}$};
        \draw[thick] (-1.5,0) -- (-2.25,-1.73*3/4) node[left]{$1^{a,+}$};
        \filldraw[thick,fill=white] (-1.5,0) circle (0.7) node{$\kappa_{-1,1,1}$};
    \end{tikzpicture} = i f^{abc} \kappa_{-1,1,1} \frac{\lbrack 12 \rbrack^3}{\lbrack 23 \rbrack \lbrack 31 \rbrack}  \quad
    && \begin{tikzpicture}[baseline={([yshift=-0.9ex]current bounding box.center)},scale=0.8]
        \draw[thick] (-1.5,0) -- (-2.25,1.73*3/4) node[left]{$2^{+,b}$};
        \draw[thick] (-1.5,0) -- (0,0) node[above]{$3^{+,c}$};
        \draw[thick] (-1.5,0) -- (-2.25,-1.73*3/4) node[left]{$1^{+,a}$};
        \filldraw[thick,fill=white] (-1.5,0) circle (0.6) node{$\kappa_{1,1,1}$};
    \end{tikzpicture} = i f^{abc} \kappa_{1,1,1} \lbrack 12 \rbrack \lbrack 23 \rbrack \lbrack 31 \rbrack \nonumber \\
    & \begin{tikzpicture}[baseline={([yshift=-0.9ex]current bounding box.center)},scale=0.8]
        \draw[thick] (-1.5,0) -- (-2.25,1.73*3/4) node[left]{$2^{+,b}$};
        \draw[thick] (-1.5,0) -- (0,0) node[above]{$3^{\phi}$};
        \draw[thick] (-1.5,0) -- (-2.25,-1.73*3/4) node[left]{$1^{+,a}$};
        \filldraw[thick,fill=white] (-1.5,0) circle (0.6) node{$\kappa_{0,1,1}$};
    \end{tikzpicture} = \sqrt{\frac{2}{N_c}} \delta^{ab}\kappa_{0,1,1} \lbrack 12 \rbrack^2 \quad
    && \begin{tikzpicture}[baseline={([yshift=-0.9ex]current bounding box.center)},scale=0.8]
        \draw[thick] (-1.5,0) -- (-2.25,1.73*3/4) node[left]{$2^{+,b}$};
        \draw[thick] (-1.5,0) -- (0,0) node[above]{$3^{\phi,c}$};
        \draw[thick] (-1.5,0) -- (-2.25,-1.73*3/4) node[left]{$1^{+,a}$};
        \filldraw[thick,fill=white] (-1.5,0) circle (0.6) node{$\kappa_{0,1,1}$};
    \end{tikzpicture} = d^{abc}\kappa_{0,1,1} \lbrack 12 \rbrack^2 \label{eq:YMvertices} \\
    & \begin{tikzpicture}[baseline={([yshift=-0.9ex]current bounding box.center)},scale=0.8]
        \draw[thick] (-1.5,0) -- (-2.25,1.73*3/4) node[left]{$2^{\phi,b}$};
        \draw[thick] (-1.5,0) -- (0,0) node[above]{$3^{\phi,c}$};
        \draw[thick] (-1.5,0) -- (-2.25,-1.73*3/4) node[left]{$1^{+,a}$};
        \filldraw[thick,fill=white] (-1.5,0) circle (0.6) node{$\kappa_{0,0,1}$};
    \end{tikzpicture} = i f^{abc} \kappa_{0,0,1}\frac{\lbrack 12 \rbrack \lbrack 31 \rbrack}{\lbrack 23 \rbrack} \nonumber
\end{align}
and their parity conjugates.  Here $f^{abc}$ are the structure constants of the gauge group and $d^{abc} = 2\, \text{Tr}[\{T^a,T^b\}T^c]$. Note that we have introduced two separate scalar fields---one adjoint and one singlet.  In a general theory, the $\kappa_{0,1,1}$ coupling of the former could be considered independent of that of the latter.  However, imposing the Jacobi identity on the OPE of soft modes requires the precise relative normalization already indicated in the two figures in the middle row above, and additionally requires
\begin{align}
\label{eq:constraint21}
        (\kappa_{-1,1,1} - \kappa_{0,0,1})\, \kappa_{0,0,1} &= 0\,, \qquad (\kappa_{-1,1,1} - \kappa_{0,0,1})\, \kappa_{0,1,1} = 0\,, \\
&\label{eq:constraint22} \kappa_{0,1,1}^2 = 2\kappa_{-1,1,1}\, \kappa_{1,1,1}\,.
\end{align}
The constraints~(\ref{eq:constraint21}) are automatically satisfied when we impose $\kappa_{-1,1,1}=\kappa_{0,0,1}$ as required by gauge invariance (at the level of a Lagrangian this identity follows when the kinetic term is written as $(D \phi)^2$ where $D$ is the gauge covariant derivative). However, the constraint~(\ref{eq:constraint22}) is novel.

Finally, we can couple the two sectors by introducing additional three-point amplitudes
\begin{equation}
\label{eq:newtwo}
    \begin{tikzpicture}[baseline={([yshift=-0.9ex]current bounding box.center)},scale=0.8]
        \draw[thick] (-1.5,0) -- (-2.25,1.73*3/4) node[left]{$2^{-,b}$};
        \draw[thick] (-1.5,0) -- (0,0) node[above]{$3^{++}$};
        \draw[thick] (-1.5,0) -- (-2.25,-1.73*3/4) node[left]{$1^{+,a}$};
        \filldraw[thick,fill=white] (-1.5,0) circle (0.7) node{$\kappa_{-1,1,2}$};
    \end{tikzpicture} = \sqrt{\frac{2}{N_c}} \delta^{ab}\kappa_{-1,1,2} \frac{\lbrack 13 \rbrack^4}{\lbrack 12 \rbrack^2} \quad
    \begin{tikzpicture}[baseline={([yshift=-0.9ex]current bounding box.center)},scale=0.8]
        \draw[thick] (-1.5,0) -- (-2.25,1.73*3/4) node[left]{$2^{+,b}$};
        \draw[thick] (-1.5,0) -- (0,0) node[above]{$3^{++}$};
        \draw[thick] (-1.5,0) -- (-2.25,-1.73*3/4) node[left]{$1^{+,a}$};
        \filldraw[thick,fill=white] (-1.5,0) circle (0.6) node{$\kappa_{1,1,2}$};
    \end{tikzpicture} = \sqrt{\frac{2}{N_c}} \delta^{ab}\kappa_{1,1,2} \lbrack 13 \rbrack^2 \lbrack 23 \rbrack^2 \\
\end{equation}
and their parity conjugates. In addition to the constraints in~(\ref{eq:constraint11}),  (\ref{eq:constraint12}), (\ref{eq:constraint21}), and~(\ref{eq:constraint22}), the celestial Jacobi identity also requires
\begin{align}
\label{eq:constraint31}
         (\kappa_{-2,2,2} - \kappa_{-1,1,2})\,\kappa_{-1,1,2} = 0 \,,\qquad &(\kappa_{-2,2,2} - \kappa_{-1,1,2})\,\kappa_{1,1,2} = 0\,, \\
         \kappa_{-1,1,2}\,\kappa_{1,1,2} = \kappa_{0,2,2} \,\kappa_{0,1,1}\,,\qquad
         &(\kappa_{0,0,2} - \kappa_{-1,1,2}) \, \kappa_{0,1,1} = 0\,, \label{eq:constraint32} \\
\label{eq:constraint33} &\hspace{-2cm}\kappa_{-1,1,2}\,\kappa_{1,1,1} = 3\kappa_{1,1,2}\,\kappa_{-1,1,1} \,.
\end{align}
Note that all of the constraints presented above are consistent with dimensional analysis since $\kappa_{s_1,s_2,s_3}$ has mass dimension $1-|s_1+s_2+s_3|$.

\section{Tree-level OPE associativity from amplitudes}
\label{sec:hardOPEassoc}

The standard way to check OPE associativity is to introduce a mode expansion of the participating operators and then compute their commutators using the OPE. If the OPE is associative, the commutator of these modes must satisfy the Jacobi identity. In the context of CCFT, the bulk interpretation of this procedure is opaque. Our goal in this paper is not just to check OPE associativity, but also to interpret the condition directly at the level of momentum space scattering amplitudes. To derive a more direct check of OPE associativity consider the following identity involving contour integrals\footnote{A similar analysis of the associativity of the one-loop OPE in self-dual Yang-Mills was performed in~\cite{Costello:2022upu}.}:
\begin{equation}
    \label{eq:contourintidentity}
    \begin{aligned}
        & \underset{\left|z_{13}\right|=2}{\oint} \dd z_1\, \underset{\left|z_{23}\right|=1}{\oint} \dd z_2\,\an{{\cal O}_{\D_1,s_1}(z_1, \zb_1) \dots {\cal O}_{\D_n,s_n}(z_n, \zb_n)}= \\
        & \underset{\left|z_{23}\right|=2}{\oint} \dd z_2\, \underset{\left|z_{13}\right|=1}{\oint} \dd z_1\,\an{{\cal O}_{\D_1,s_1}(z_1, \zb_1) \dots {\cal O}_{\D_n,s_n}(z_n, \zb_n)}
        +\underset{\left|z_{23}\right|=2}{\oint} \dd z_2\, \underset{\left|z_{12}\right|=1}{\oint} \dd z_1\,\an{{\cal O}_{\D_1,s_1}(z_1, \zb_1) \dots {\cal O}_{\D_n,s_n}(z_n, \zb_n)}.
    \end{aligned}
\end{equation}
This is valid as long as all of the ($\Delta_i,s_i$) are such that the correlation functions are single-valued functions of $z_i$. Practically, we evaluate these integrals using the OPE. Thus the identity serves as a check on the OPE associativity. We can write this in a condensed form as
\begin{align}
\label{eq:doubleresidue1}
  \left[\underset{2\to 3}{\text{Res}}\, \underset{1\to 2}{\text{Res}} - \underset{1\to 3}{\text{Res}}\, \underset{2\to 3}{\text{Res}}+\underset{2\to 3}{\text{Res}}\,\underset{1\to 3}{\text{Res}}\right]\an{{\cal O}_{\D_1,s_1}(z_1, \zb_1) \dots {\cal O}_{\D_n,s_n}(z_n, \zb_n)} = 0\,.
\end{align}
The correlator in~(\ref{eq:doubleresidue1}) is nothing but the Mellin transform of the amplitude. If we parameterize the momenta of outgoing massless external particles as
\begin{align}
    p^{\mu}_i \sim \lam_i \lamt_i \qquad \text{with } \qquad  \lam_i = \sqrt{2\omega_i}\begin{pmatrix} 1 \\ z_i\end{pmatrix} \qquad \text{and } \qquad \lamt_i = \sqrt{2\omega_i}\begin{pmatrix} 1 \\ \zb_i\end{pmatrix},
\end{align}
the transformation from momentum to boost eigenstates is implemented via the Mellin transform
\begin{align}
    \label{eq:Mellintransform}
    \an{{\cal O}_{\D_1,s_1}\left(z_1, \zb_1\right) \dots {\cal O}_{\D_n,s_n}\left(z_n, \zb_n\right)} = \int \prod_{i=1}^n \frac{\dd \omega_i}{\omega_i^{1-\D_i}} \mA_n\left(\left\lbrace \lam_1, \lamt_1\right\rbrace^{s_1}, \left\lbrace \lam_2, \lamt_2\right\rbrace^{s_2}, \dots, \left\lbrace \lam_n, \lamt_n\right\rbrace^{s_n}\right).
\end{align}
Demanding
\begin{align}
    \label{eq:doubleresideuamp1}
    \left[\underset{2\to 3}{\text{Res}}\, \underset{1\to 2}{\text{Res}} - \underset{1\to 3}{\text{Res}}\, \underset{2\to 3}{\text{Res}}+\underset{2\to 3}{\text{Res}}\,\underset{1\to 3}{\text{Res}}\right] \,\mA_n\left(\left\lbrace \lam_1, \lamt_1\right\rbrace^{s_1}, \left\lbrace \lam_2, \lamt_2\right\rbrace^{s_2}, \dots, \left\lbrace \lam_n, \lamt_n\right\rbrace^{s_n}\right)  = 0
\end{align}
ensures that the same is true of the correlator. The upshot is that we can check associativity of the celestial OPE by evaluating double residues directly on the amplitude rather than dealing with the celestial correlators.

The first step is to isolate the collinear limits. Bearing in mind that we will later be interested in examining the effects of~(\ref{eq:doubleresideuamp1}) on amplitudes with various higher derivative corrections, we will analyze the collinear limit using the all-line shift recursion relations~\cite{Cachazo:2004kj,Elvang:2008vz}, reviewed briefly in Appendix~\ref{sec:CSW}. We begin by using~(\ref{eq:CSWrecursion}) to isolate the collinear channel as the momentum of particle one approaches that of particle two
\begin{align}
&{\cal A}_n \left(\left\lbrace \lam_1, \lamt_1\right\rbrace^{s_1}, \left\lbrace \lam_2, \lamt_2\right\rbrace^{s_2}, \dots, \left\lbrace \lam_n, \lamt_n\right\rbrace^{s_n}\right) \\
\nonumber=& \sum_{s_I} \hat{\mA}_3  \left(\left\lbrace \lamh_1, \lamt_1\right\rbrace^{s_1}, \left\lbrace \lamh_2, \lamt_2\right\rbrace^{s_2},\left\lbrace \lamh_I, \hat{\lamt}_I\right\rbrace^{-s_I}\right)  \frac{1}{\an{12}\sq{12}} \hat{\mA}_{n-1} \left(\left\lbrace \lamh_I, \hat{\lamt}_I\right\rbrace^{s_I}, \dots, \left\lbrace \lamh_n, \lamt_n\right\rbrace^{s_n}\right)\\
&\nonumber \qquad\qquad + \left\lbrace\text{other channels}\right\rbrace.
\end{align}
In terms of these, the residue is
\begin{align}
\label{eq:residue12}
 \underset{z_1 \to z_2}{\text{Res}} \, \mA_n &=    \sum_{s_I} \mA_3  \left(\lamt_1^{s_1}, \lamt_2^{s_2}, \lamt_I^{-s_I}\right)  \frac{1}{2\sqrt{\omega_1\omega_2}\sq{12}} {\cal A}_{n-1} \left(\left\lbrace \lam_I, \lamt_I\right\rbrace^{s_I}, \dots, \left\lbrace \lam_n, \lamt_n\right\rbrace^{s_n}\right)
\end{align}
where, on the right-hand side, in the limit $z_{12} = 0$ we have
\begin{align}
    \lam_1 = \sqrt{\frac{\omega_1}{\omega_1+\omega_2}}\lam_I\,, \qquad \lam_2 = \sqrt{\frac{\omega_2}{\omega_1+\omega_2}} \lam_I\,, \qquad \lamt_I =\sqrt{\frac{\omega_1}{\omega_1+\omega_2}} \lamt_1 + \sqrt{\frac{\omega_2}{\omega_1+\omega_2}}\lamt_2\,.
\end{align}
In arriving at~(\ref{eq:residue12}), we have made use of the fact that only anti-holomorphic three-point amplitudes contribute to the collinear limit (see Appendix~\ref{sec:CSW}) and that the terms from other channels drop out in this limit. Furthermore, the subamplitude $\mA_{n-1}$ now depends only on the unshifted momenta as the deformation parameter $\alpha$ vanishes in the collinear limit (see~(\ref{eq:collinearCSW})). Applying this formula a second time, to compute the double residue, therefore translates~(\ref{eq:doubleresideuamp1}) into
\begin{align}
    \nonumber
    \sum_{s_{I_1}}&\left[ \frac{1}{\zb_{12}\zb_{I_1 3}}\frac{1}{\omega_1+\omega_2} \mA_3 \left(\lamt_1^{s_1}, \lamt_2^{s_2}, \lamt_{I_1}^{-s_{I_1}}\right)\mA_3 \left(\lamt_{I_1}^{s_{I_1}}, \lamt_3^{s_3}, \lamt_{I_2}^{-s_{I_2}}\right)\right.\\
    &\left.\label{eq:doubleresidueamp2}+\frac{1}{\zb_{23}\zb_{I_1 1}} \frac{1}{\omega_2+\omega_3} \mA_3 \left(\lamt_2^{s_2}, \lamt_3^{s_3}, \lamt_{I_1}^{-s_{I_1}}\right)\mA_3 \left(\lamt_{I_1}^{s_{I_1}}, \lamt_1^{s_1}, \lamt_{I_2}^{-s_{I_2}}\right)\right.\\
    &\left.+\frac{1}{\zb_{31}\zb_{I_1 2}} \frac{1}{\omega_1+\omega_3} \mA_3 \left(\lamt_3^{s_3}, \lamt_1^{s_1}, \lamt_{I_1}^{-s_{I_1}}\right)\mA_3 \left(\lamt_{I_1}^{s_{I_1}}, \lamt_2^{s_2}, \lamt_{I_2}^{-s_{I_2}}\right)\right] = 0\, \nonumber
\end{align}
with $\bar{z}_{I_1} = \frac{\omega_1 \bar{z}_1 + \omega_2 \bar{z}_2}{\omega_1 + \omega_2}$. (We emphasize that this equation must hold for arbitrary $s_{I_2}$.) Using the three-point amplitudes defined in the previous section, it is now a straightforward (if slightly tedious) exercise to show that~(\ref{eq:doubleresidueamp2}) is true only when the constraints~(\ref{eq:constraint11}), (\ref{eq:constraint12}), (\ref{eq:constraint21}), (\ref{eq:constraint22}) and~(\ref{eq:constraint31})--(\ref{eq:constraint33}) are satisfied.

\section{Amplitudes in EFTs with celestial dual}
\label{sec:implications}

In this section, we will examine the properties of amplitudes in effective theories that satisfy~(\ref{eq:doubleresidueamp2}) or equivalently, the constraints reviewed in Section~\ref{sec:softOPEassoc}.

\subsection{Four-point amplitudes in the graviton-scalar sector}
\label{sec:4ptgrav}

We start with amplitudes involving external gravitons, which were computed in~\cite{Cohen:2010mi,Broedel:2012rc}. The amplitude involving four positive helicity gravitons is\footnote{We will omit explicitly displaying the dependence of the amplitude on $\lam, \lamt$ for brevity and only display the helicities.}
\begin{align}
\label{diagramone}
         \mA_{4}\left( 1^{++}, 2^{++}, 3^{++}, 4^{++} \right)
        =  &\begin{tikzpicture}[baseline={([yshift=-0.9ex]current bounding box.center)},scale=0.9]
                \draw[thick] (-1.1,0) node[above right]{$\pm\pm$} -- (0,0) -- (1.1,0) node[above left]{$\mp\mp$};
                \draw[thick] (-1.5,0) -- (-2,1.73/2) node[left]{$2^{++}$};
                \draw[thick] (1.5,0) -- (2,1.73/2) node[right]{$3^{++}$};
                \draw[thick] (-1.5,0) -- (-2,-1.73/2) node[left]{$1^{++}$};
                \draw[thick] (1.5,0) -- (2,-1.73/2) node[right]{$4^{++}$};
                \filldraw[thick,fill=white] (-1.5,0) circle (0.4);
                \filldraw[thick,fill=white] (1.5,0) circle (0.4);
            \end{tikzpicture} + \begin{tikzpicture}[baseline={([yshift=-0.9ex]current bounding box.center)},scale=0.9]
                \draw[thick] (-1.1,0) node[above right]{$\phi$} -- (0,0) -- (1.1,0) node[above left]{$\phi$};
                \draw[thick] (-1.5,0) -- (-2,1.73/2) node[left]{$2^{++}$};
                \draw[thick] (1.5,0) -- (2,1.73/2) node[right]{$3^{++}$};
                \draw[thick] (-1.5,0) -- (-2,-1.73/2) node[left]{$1^{++}$};
                \draw[thick] (1.5,0) -- (2,-1.73/2) node[right]{$4^{++}$};
                \filldraw[thick,fill=white] (-1.5,0) circle (0.4);
                \filldraw[thick,fill=white] (1.5,0) circle (0.4);
            \end{tikzpicture}\\
            &\nonumber \qquad +\  \text{(cyclic of 1,2,3)}
\end{align}
where $\text{(cyclic of 1,2,3)}$ refers to the diagrams obtained by making the stated replacements. Evaluating the diagrams results in
\begin{align}
    \mA_{4}\left( 1^{++}, 2^{++}, 3^{++}, 4^{++} \right)
    = & \, \kappa_{-2,2,2}\kappa_{2,2,2}\left( \frac{\lbrack 12\rbrack^{5}\lbrack 34\rbrack^{2}}{\langle 12 \rangle} \frac{\langle 1 X \rangle^{2} \langle 2 X \rangle^{2}}{\langle 3 X \rangle^{2} \langle 4 X \rangle^{2}} + \frac{\lbrack 34\rbrack^{5}\lbrack 12\rbrack^{2}}{\langle 34 \rangle} \frac{\langle 3 X \rangle^{2} \langle 4 X \rangle^{2}}{\langle 1 X \rangle^{2} \langle 2 X \rangle^{2}} \right)\nonumber \\
    &+ \kappa_{0,2,2}^2 \frac{\lbrack 12\rbrack^{4}\lbrack 34\rbrack^{4}}{s_{12}}  +\ \text{(cyclic of 1,2,3)} \nonumber\\
    = &  \, \left(10\kappa_{-2,2,2}\kappa_{2,2,2}-3\kappa_{0,2,2}^2\right)\, s_{12} s_{13} s_{23} \frac{\lbrack 12\rbrack\lbrack 23\rbrack\lbrack 34\rbrack\lbrack 41\rbrack}{\langle 12 \rangle \langle 23 \rangle \langle 34 \rangle \langle 41 \rangle } \\
    = &\, 0 \nonumber
\end{align}
where we used $s_{ij} = \an{ij}\sq{ij}$. In the last line, we have imposed the constraint in~(\ref{eq:constraint12}) to conclude that the amplitude vanishes.

The remaining four-graviton amplitudes $\mA \left(1^{++},2^{++},3^{++},4^{--} \right)$, $\mA\left(1^{++},2^{--},3^{--},4^{++}\right)$, and their parity conjugates are not all-line shift constructible since they violate the condition~(\ref{eq:CSWconstructability}). Instead, they were computed in~\cite{Cohen:2010mi,Broedel:2012rc, Bai:2016hui}:
\begin{equation}
\label{eq:nonvanishing4ptgravity}
\begin{aligned}
    &\mA \left(1^{++},2^{++},3^{++},4^{--} \right) = \kappa_{2,2,2} \kappa_{-2,-2,2} (\langle 14 \rangle \lbrack 13\rbrack \langle 34 \rangle)^{2} \frac{\lbrack 12\rbrack\lbrack 23\rbrack\lbrack 31\rbrack}{\langle 12\rangle \langle 23 \rangle \langle 31 \rangle} \,,\\
     &\mA \left(1^{++},2^{--},3^{--},4^{++}\right) = \frac{\left(\an{23}\sq{14}\right)^4}{s_{14}} \left(\kappa_{2,2,2}^2 s_{12}\, s_{13} - \kappa_{2,2,0}^2 + \kappa^2_{2,2,-2} \frac{1}{s_{12}\, s_{13}}\right) \,,
\end{aligned}
\end{equation}
and we note that these amplitudes are non-vanishing, even on the support of the associativity constraints.

Moving on to amplitudes involving external scalars in addition to gravitons, the only amplitude that is all-line shift constructible is
\begin{align}
   \mA_4 \left(1^{++}, 2^{++}, 3^{++}, 4^{\phi}\right) =
   &
   \begin{tikzpicture}[baseline={([yshift=-0.9ex]current bounding box.center)},scale=0.9]
        \draw[thick] (-1.1,0) node[above right]{$--$} -- (0,0) -- (1.1,0) node[above left]{$++$};
        \draw[thick] (-1.5,0) -- (-2,1.73/2) node[left]{$2^{++}$};
        \draw[thick] (1.5,0) -- (2,1.73/2) node[right]{$3^{++}$};
        \draw[thick] (-1.5,0) -- (-2,-1.73/2) node[left]{$1^{++}$};
        \draw[thick] (1.5,0) -- (2,-1.73/2) node[right]{$4^\phi$};
        \filldraw[thick,fill=white] (-1.5,0) circle (0.4);
        \filldraw[thick,fill=white] (1.5,0) circle (0.4);
    \end{tikzpicture}
    +
    \begin{tikzpicture}[baseline={([yshift=-0.9ex]current bounding box.center)},scale=0.9]
        \draw[thick] (-1.1,0) node[above right]{$\phi$} -- (0,0) -- (1.1,0) node[above left]{$\phi$};
        \draw[thick] (-1.5,0) -- (-2,1.73/2) node[left]{$2^{++}$};
        \draw[thick] (1.5,0) -- (2,1.73/2) node[right]{$3^{++}$};
        \draw[thick] (-1.5,0) -- (-2,-1.73/2) node[left]{$1^{++}$};
        \draw[thick] (1.5,0) -- (2,-1.73/2) node[right]{$4^\phi$};
        \filldraw[thick,fill=white] (-1.5,0) circle (0.4);
        \filldraw[thick,fill=white] (1.5,0) circle (0.4);
    \end{tikzpicture}\nonumber\\
    \nonumber & \qquad + \ \text{(cyclic of 1,2,3)}\\
    =  &\,\kappa_{-2,2,2} \kappa_{2,2,0} \frac{\sq{12}\sq{34}^4}{\an{12}} \frac{\an{4X}^4}{\an{1X}^2\an{2X}^2} + \kappa_{2,2,0}\kappa_{0,0,2} \frac{\sq{12}^3\sq{34}^2}{\an{12}} \frac{\an{4X}^2}{\an{3X}^2}\\
    \nonumber & \qquad +\ \text{(cyclic of 1,2,3)}\\
    =  &\, 0\nonumber
\end{align}
and it vanishes on the support of the constraints~(\ref{eq:constraint11}).

\subsection{Four-point amplitudes in the gluon-scalar sector}
\label{sec:4ptgluonscalar}

The situation in the gluon-scalar sector is similar: the constraints lead to the vanishing of all-line shift constructible amplitudes. In this sector the only all-line shift constructible all-gluon amplitude is
\begin{multline}
 \mA_4\left(1^{+,a_1}, 2^{+,a_2}, 3^{+,a_3}, 4^{+,a_4}\right) =
    \begin{tikzpicture}[baseline={([yshift=-0.9ex]current bounding box.center)},scale=0.8]
        \draw[thick] (-1.1,0) node[above right]{$\mp$} -- (0,0) -- (1.1,0) node[above left]{$\pm$};
        \draw[thick] (-1.5,0) -- (-2,1.73/2) node[left]{$2^{+,a_2}$};
        \draw[thick] (1.5,0) -- (2,1.73/2) node[right]{$3^{+,a_3}$};
        \draw[thick] (-1.5,0) -- (-2,-1.73/2) node[left]{$1^{+,a_1}$};
        \draw[thick] (1.5,0) -- (2,-1.73/2) node[right]{$4^{+,a_4}$};
        \filldraw[thick,fill=white] (-1.5,0) circle (0.4);
        \filldraw[thick,fill=white] (1.5,0) circle (0.4);
    \end{tikzpicture}\\
    +
    \begin{tikzpicture}[baseline={([yshift=-0.9ex]current bounding box.center)},scale=0.8]
        \draw[thick] (-1.1,0) node[above right]{$\phi^{b}$} -- (0,0) -- (1.1,0) node[above left]{$\phi^{b}$};
        \draw[thick] (-1.5,0) -- (-2,1.73/2) node[left]{$2^{+,a_2}$};
        \draw[thick] (1.5,0) -- (2,1.73/2) node[right]{$3^{+,a_3}$};
        \draw[thick] (-1.5,0) -- (-2,-1.73/2) node[left]{$1^{+,a_1}$};
        \draw[thick] (1.5,0) -- (2,-1.73/2) node[right]{$4^{+,a_4}$};
        \filldraw[thick,fill=white] (-1.5,0) circle (0.4);
        \filldraw[thick,fill=white] (1.5,0) circle (0.4);
    \end{tikzpicture}
    +
    \begin{tikzpicture}[baseline={([yshift=-0.9ex]current bounding box.center)},scale=0.8]
        \draw[thick] (-1.1,0) node[above right]{$\phi$} -- (0,0) -- (1.1,0) node[above left]{$\phi$};
        \draw[thick] (-1.5,0) -- (-2,1.73/2) node[left]{$2^{+,a_2}$};
        \draw[thick] (1.5,0) -- (2,1.73/2) node[right]{$3^{+,a_3}$};
        \draw[thick] (-1.5,0) -- (-2,-1.73/2) node[left]{$1^{+,a_1}$};
        \draw[thick] (1.5,0) -- (2,-1.73/2) node[right]{$4^{+,a_4}$};
        \filldraw[thick,fill=white] (-1.5,0) circle (0.4);
        \filldraw[thick,fill=white] (1.5,0) circle (0.4);
    \end{tikzpicture}
    + \quad \text{(cyclic of 1,2,3)}
\end{multline}
A straightforward calculation yields
\begin{align}\label{eq:4+F3}
    \nonumber & \mA_4\left(1^{+,a_1}, 2^{+,a_2}, 3^{+,a_3}, 4^{+,a_4}\right)  \\
    \nonumber = & -f^{a_1a_2e}f^{ea_3a_4} \, \kappa_{-1,1,1}\kappa_{1,1,1} \left( \frac{\lbrack 12\rbrack^{2}\lbrack 34\rbrack}{\langle 12 \rangle} \frac{\langle 1 X \rangle \langle 2 X \rangle}{\langle 3 X \rangle \langle 4 X \rangle} + \frac{\lbrack 34\rbrack^{2}\lbrack 12\rbrack}{\langle 34 \rangle} \frac{\langle 3 X \rangle \langle 4 X \rangle}{\langle 1 X \rangle \langle 2 X \rangle} \right) \\
    & +\frac{2 \kappa_{0,1,1}^2}{N_c} \left( \delta^{a_1 a_2} \delta^{a_3 a_4} \frac{\lbrack 12 \rbrack^2 \lbrack 34 \rbrack ^2}{\an{12}\sq{12}} \right) +\kappa_{0,1,1}^2 \left( d^{a_1 a_2 b} d^{a_3 a_4 b} \frac{\lbrack 12 \rbrack^2 \lbrack 34 \rbrack ^2}{\an{12}\sq{12}} \right) + (\text{cyclic of 1,2,3}) \\
    = & \left( 4 \kappa_{-1,1,1} \kappa_{1,1,1} - 2\kappa_{0,1,1}^2 \right) \frac{\lbrack 13 \rbrack \lbrack 24 \rbrack^2}{\langle 13 \rangle} \text{Tr}\lbrack T^{a_1} T^{a_2} T^{a_3} T^{a_4} \rbrack + ( \text{permutation of 1,2,3} ) \nonumber\,.
\end{align}
Here we used properties of $f^{abc}$ and $d^{abc}$, and in the last line the constraint $\kappa_{0,1,1}^2 = 2\kappa_{-1,1,1}\kappa_{1,1,1}$ from~(\ref{eq:constraint22}) to find that this amplitude vanishes as well. The remaining four-gluon amplitudes are not all-line shift constructible and are non-vanishing for generic couplings, even when all the constraints from Section~\ref{sec:softOPEassoc} are imposed.

We also find that the four-point amplitude involving three positive helicity gluons and one scalar (either the adjoint or singlet) is  vanishing. For the adjoint scalar, the full amplitude reads (here we omit on each line ``+ cyclic(1,2,3)'')
\begin{align}
    &\mA_4\left(1^{+,a_1},2^{+,a_2},3^{+,a_3},4^{\phi,a_4}\right) \nonumber\\
    =\, & \nonumber\begin{tikzpicture}[baseline={([yshift=-0.9ex]current bounding box.center)},scale=0.8]
        \draw[thick] (-1.1,0) node[above right]{$-$} -- (0,0) -- (1.1,0) node[above left]{$+$};
        \draw[thick] (-1.5,0) -- (-2,1.73/2) node[left]{$2^{+,a_2}$};
        \draw[thick] (1.5,0) -- (2,1.73/2) node[right]{$3^{+,a_3}$};
        \draw[thick] (-1.5,0) -- (-2,-1.73/2) node[left]{$1^{+,a_1}$};
        \draw[thick] (1.5,0) -- (2,-1.73/2) node[right]{$4^{\phi,a_4}$};
        \filldraw[thick,fill=white] (-1.5,0) circle (0.4);
        \filldraw[thick,fill=white] (1.5,0) circle (0.4);
    \end{tikzpicture}
    +
    \begin{tikzpicture}[baseline={([yshift=-0.9ex]current bounding box.center)},scale=0.8]
        \draw[thick] (-1.1,0) node[above right]{$\phi^{b}$} -- (0,0) -- (1.1,0) node[above left]{$\phi^{b}$};
        \draw[thick] (-1.5,0) -- (-2,1.73/2) node[left]{$2^{+,a_2}$};
        \draw[thick] (1.5,0) -- (2,1.73/2) node[right]{$3^{+,a_3}$};
        \draw[thick] (-1.5,0) -- (-2,-1.73/2) node[left]{$1^{+,a_1}$};
        \draw[thick] (1.5,0) -- (2,-1.73/2) node[right]{$4^{\phi,a_4}$};
        \filldraw[thick,fill=white] (-1.5,0) circle (0.4);
        \filldraw[thick,fill=white] (1.5,0) circle (0.4);
    \end{tikzpicture}\nonumber\\
    =\, & i f^{a_1 a_2 b} d^{a_3 a_4 b} \kappa_{-1,1,1} \kappa_{0,1,1} \frac{\lbrack 34 \rbrack^2}{\langle 12 \rangle} \frac{\langle 4 X \rangle^2}{\langle 1 X \rangle \langle 2 X \rangle} + i d^{a_1 a_2 b} f^{a_3 a_4 b} \kappa_{0,1,1} \kappa_{0,0,1} \frac{\lbrack 12 \rbrack \lbrack 34 \rbrack}{\langle 12 \rangle} \frac{\langle 4 X \rangle}{\langle 3 X \rangle}   \\
    =\, & i f^{a_1 a_2 b} d^{a_3 a_4 b} \kappa_{-1,1,1} \kappa_{0,1,1} \frac{\lbrack 34 \rbrack^2}{\langle 12 \rangle} \frac{\langle 4 X \rangle^2}{\langle 1 X \rangle \langle 2 X \rangle} + \left( i f^{a_2 a_3 b} d^{a_1 a_4 b} - i f^{a_3 a_1 b} d^{a_2 a_4 b} \right) \kappa_{0,1,1} \kappa_{0,0,1} \frac{\lbrack 12 \rbrack \lbrack 34 \rbrack}{\langle 12 \rangle} \frac{\langle 4 X \rangle}{\langle 3 X \rangle} \nonumber \\
    =\, & i f^{a_1 a_2 b} d^{a_3 a_4 b} \left( \kappa_{-1,1,1} \kappa_{0,1,1} \frac{\lbrack 34 \rbrack^2 \langle 4 X \rangle^2}{\langle 12 \rangle \langle 1 X \rangle \langle 2 X \rangle} - \kappa_{0,1,1} \kappa_{0,0,1} \frac{\lbrack 23 \rbrack \lbrack 14 \rbrack \langle 4 X \rangle}{\langle 23 \rangle \langle 1 X \rangle} + \kappa_{0,1,1} \kappa_{0,0,1} \frac{\lbrack 31 \rbrack \lbrack 24 \rbrack \langle 4 X \rangle}{\langle 31 \rangle \langle 2 X \rangle} \right) \nonumber\\
    =\, & i f^{a_1 a_2 b} d^{a_3 a_4 b} \left( \kappa_{-1,1,1} \kappa_{0,1,1} - \kappa_{0,1,1} \kappa_{0,0,1} \right) \frac{\lbrack 34 \rbrack^2 \langle 4 X \rangle^2}{\langle 12 \rangle \langle 1 X \rangle \langle 2 X \rangle}  \nonumber \\
    =\, & 0  \nonumber
\end{align}
on the support of the constraints~(\ref{eq:constraint31}), while the amplitude for the singlet scalar is:
\begin{align}
    &\mA_4\left(1^{+,a_1},2^{+,a_2},3^{+,a_3},4^{\phi}\right)
    =\,  \nonumber\begin{tikzpicture}[baseline={([yshift=-0.9ex]current bounding box.center)},scale=0.8]
        \draw[thick] (-1.1,0) node[above right]{$-$} -- (0,0) -- (1.1,0) node[above left]{$+$};
        \draw[thick] (-1.5,0) -- (-2,1.73/2) node[left]{$2^{+,a_2}$};
        \draw[thick] (1.5,0) -- (2,1.73/2) node[right]{$3^{+,a_3}$};
        \draw[thick] (-1.5,0) -- (-2,-1.73/2) node[left]{$1^{+,a_1}$};
        \draw[thick] (1.5,0) -- (2,-1.73/2) node[right]{$4^{\phi}$};
        \filldraw[thick,fill=white] (-1.5,0) circle (0.4);
        \filldraw[thick,fill=white] (1.5,0) circle (0.4);
    \end{tikzpicture} \quad + \quad (\text{cyclic of 1,2,3}) \\
     &\qquad =\,  i \sqrt{\frac{2}{N_c}} \kappa_{-1,1,1} \kappa_{0,1,1} f^{a_1 a_2 b} \delta^{a_3 b} \frac{\lbrack 34 \rbrack^2}{\langle 12 \rangle} \frac{\langle 4 X \rangle^2}{\langle 1 X \rangle \langle 2 X \rangle} \quad + \quad (\text{cyclic of 1,2,3})\\
     &\qquad =\,  i \sqrt{\frac{2}{N_c}} \kappa_{-1,1,1} \kappa_{0,1,1} f^{a_1 a_2 a_3} \left( \frac{\lbrack 34 \rbrack^2 \langle 4 X \rangle^2}{\langle 12 \rangle \langle 1 X \rangle \langle 2 X \rangle} + \frac{\lbrack 14 \rbrack^2 \langle 4 X \rangle^2}{\langle 23 \rangle \langle 2 X \rangle \langle 3 X \rangle} + \frac{\lbrack 24 \rbrack^2 \langle 4 X \rangle^2}{\langle 31 \rangle \langle 3 X \rangle \langle 1 X \rangle} \right) \nonumber\\
     &\qquad =\,  0 \,. \nonumber
\end{align}

\subsection{Amplitudes of arbitrary multiplicity}
\label{sec:npt}

We can now generalize the results of the previous sections to amplitudes of arbitrary multiplicity.

For amplitudes that are all-line shift constructible, the general statement is that all contributions that involve purely holomorphic (or purely anti-holomorphic) vertices to amplitudes must vanish due to the constraints. So, in particular, if an amplitude is all-line shift constructible and each term breaks down into only purely holomorphic or purely anti-holomorphic three-point building blocks, then the entire amplitude must vanish.

To see this, let us start by constructing an arbitrary four-point amplitude by using the all-line shift recursion relations to glue together two anti-holomorphic three-point amplitudes:
\begin{align}
 &\mA_4 \left(\left\lbrace \lam_1, \lamt_1\right\rbrace^{s_1}, \left\lbrace \lam_2, \lamt_2\right\rbrace^{s_2}, \left\lbrace \lam_3, \lamt_3\right\rbrace^{s_3}, \left\lbrace \lam_4, \lamt_4\right\rbrace^{s_4}\right)  \nonumber  \\  &\qquad\qquad= \begin{tikzpicture}[baseline={([yshift=-1ex]current bounding box.center)}]
            \draw[thick] (0,0) -- (120:1) node[left]{$2^{s_2}$};
            \draw[thick] (0,0) -- (240:1) node[left]{$1^{s_1}$};
            \draw[thick] (0,0) node[above right]{$\ P_{I}^{-s_I}$} -- (0:3) node[above left]{$-P_{I}^{s_I}\ \,$} -- ++ (60:1) node[right]{$3^{s_3}$} (0:3) -- ++ (-60:1) node[right]{$4^{s_4}$};
            \filldraw[thick,fill=white] (0,0) circle (.3);
            \filldraw[thick,fill=white] (3,0) circle (.3);
        \end{tikzpicture} + \quad \text{(cyclic of 1,2,3)} \label{eq:diag1234}\\
&\nonumber\qquad\qquad= \sum_{s_I} \left[\mA_3  \left(\lamt_1^{s_1}, \lamt_2^{s_2}, \hat{\lamt}_I^{-s_I}\right)  \frac{1}{\an{12}\sq{12}} \mA_{3} \left(\hat{\lamt}_I^{s_I},\lamt_3^{s_3}, \lamt_4^{s_4}\right) + \text{(cyclic of 1,2,3)} \right].
\end{align}
Using momentum conservation, we can rewrite this as
\begin{align}
 \nonumber  &\frac{\sq{34}}{\an{12}}\sum_{s_I} \left[\mA_3  \left(\lamt_1^{s_1}, \lamt_2^{s_2}, \hat{\lamt}_I^{-s_I}\right)  \frac{1}{\sq{12}\sq{34}} \mA_{3} \left(\hat{\lamt}_I^{s_I},\lamt_3^{s_3}, \lamt_4^{s_4}\right)\right.\\
     &\qquad\qquad+\mA_3  \left(\lamt_1^{s_1}, \lamt_3^{s_3}, \hat{\lamt}_I^{-s_I}\right)  \frac{1}{\sq{31}\sq{24}} \mA_{3} \left(\hat{\lamt}_I^{s_I},\lamt_2^{s_2}, \lamt_4^{s_4}\right)\\
&\nonumber\left.\qquad\qquad+\mA_3  \left(\lamt_1^{s_1}, \lamt_2^{s_4}, \hat{\lamt}_I^{-s_I}\right)  \frac{1}{\sq{14}\sq{23}} \mA_{3} \left(\hat{\lamt}_I^{s_I},\lamt_3^{s_3}, \lamt_4^{s_2}\right)\right],
\end{align}
which must hold for arbitrary $s_4$. This is precisely the double residue condition~(\ref{eq:doubleresidueamp2}) on the support of four-point momentum conservation! Hence we conclude that tree-level associativity of the celestial OPE forces any four-point amplitude constructed solely from anti-holomorphic vertices to vanish\footnote{The apparent asymmetry between holomorphic and anti-holomorphic in our discussion arises from our choice of using the holomorphic all-line shift in Appendix~\ref{sec:CSW}. Of course, all conclusions hold for parity conjugate amplitudes as well, and could be manifested by using an anti-holomorphic all-line shift instead. In constructing higher-point amplitudes recursively, one has the freedom to independently choose the holomorphic or anti-holomorphic shift term-by-term and in various levels of the recursion.}.

The construction of higher-point amplitudes proceeds in a similar manner. However, unlike the four-particle case, we can encounter amplitudes that are all-line shift constructible but non-vanishing. This can occur because none of the non-zero four-point amplitudes are purely anti-holomorphic or purely holomorphic, so any higher-point amplitudes (when constructed via all-line shift recursion) that receive contributions from these amplitudes will also be non-vanishing. An example is the five-point all-plus amplitude $\mA_5(1^+,2^+,3^+,4^+,5^+)$, which is all-line shift constructible but has a term involving the product of two non-zero amplitudes $\mA_3(1^+,2^+,3^+)\times \mA_4(1^-,2^+,3^+,4^+)$, the second of which includes both holomorphic and anti-holomorphic vertices. However, constructible higher-point amplitudes will be vanishing at tree level if they do not receive contributions from any channels other than those involving only purely holomorphic or anti-holomorphic vertices as building blocks.

\section{Discussion}

The constraints~(\ref{eq:constraint11}), (\ref{eq:constraint12}), (\ref{eq:constraint21}), (\ref{eq:constraint22}) and~(\ref{eq:constraint31})--(\ref{eq:constraint33}) are incredibly restrictive. Most apparently sensible theories fail to satisfy them. Heterotic string theory (compactified on a torus to 4D) ostensibly fails to satisfy them due to the presence of $R^2 \phi$ terms and the absence of an $R^3$ term.\footnote{We thank Mina Himwich for pointing this out.} However, the four-positive helicity graviton amplitude vanishes in this theory as can be seen from the double copy construction~\cite{Azevedo:2018dgo}. {This is because $\kappa_{2,2,2}$ vanishes and the two scalars (axion and dilaton) give contributions to the left-hand side of~(\ref{eq:constraint12}) that vanish when summed. While the vanishing of the four-positive helicity amplitude does not guarantee associativity, it serves as an easy (particularly when the spectrum involves multiple particles with the same helicity) but powerful check on OPE associativity. It is intriguing to wonder about other  contributions (massive particles, bound states, resonances, extended objects) to the OPE of massless particles, see for e.g.~\cite{Garcia-Sepulveda:2022lga}. If some or all of these contributions are indeed present, this suggests that the OPE in~(\ref{eq:genericOPE}) is incomplete and these additional contributions might modify the analysis in this paper.  Alternatively, theories which fail to satisfy these constraints simply do not have celestial duals.}

Interestingly, one {other} example of a theory which does satisfy the constraints is the chiral higher-spin theory studied in~\cite{Metsaev:1991nb,Metsaev:1991mt,Ponomarev:2016lrm}. Working in the light-cone approach, they derived the following solution for the coupling constants after requiring Poincar{\'e} symmetry:
\begin{equation}
    \kappa_{s_1,s_2,s_3} \sim \frac{(l_P)^{s_1+s_2+s_3-1}}{\Gamma(s_1+s_2+s_3)}
    \qquad s_1+s_2+s_3>0\,,
\label{eq:higherspin}
\end{equation}
where $l_P$ is a parameter with dimension of length. One can easily check that~(\ref{eq:higherspin}) satisfies all the constraints in (\ref{eq:constraint11}), (\ref{eq:constraint12}), (\ref{eq:constraint21}), (\ref{eq:constraint22}) and~(\ref{eq:constraint31})--(\ref{eq:constraint33}).

In light of~\cite{Ball:2021tmb,Costello:2022upu} it is natural to speculate about the prospect of a self-dual theory on the support of the associativity constraints, generalizing self-dual Yang-Mills and self-dual gravity with some higher derivative corrections. The key property of self-dual Yang-Mills and self-dual gravity is that only the anti-holomorphic three-point vertices are nonzero, and all higher-point amplitudes vanish at tree level. After including the higher-derivative interactions, if we continue to keep only the anti-holomorphic three-point vertices and set the holomorphic three-point vertices to zero, then all higher-point tree-level amplitudes will be vanishing only when the constraints are satisfied as we have shown. We leave the study of such possibilities for future work.

\acknowledgments

We are grateful to the organizers and participants of the workshop ``Possible and Impossible in Effective Field Theory: From S-Matrix to the Swampland'', where part of this work was presented, for useful comments and discussion. We also thank Rishabh Bhardwaj, Shounak De, Mina Himwich, Yangrui Hu, Konstantinos Koutrolikos, Luke Lippstreu, Xianlong Liu, Jorge Mago, Oliver Schlotterer and Andy Strominger for stimulating comments and correspondence. This work was supported in part by the US Department of Energy under contract {DE}-{SC}0010010 Task F and by Simons Investigator Award \#376208 (AV).

\appendix

\section{All-line shift recursion relations}
\label{sec:CSW}

In this appendix we will review the all-line shift recursion relations following~\cite{Cohen:2010mi}.

The all-line shift recursion relations are based on shifting all of the external momenta. For the purpose this appendix, we treat $\lam, \lamt$ (consequently $z, \zb$) as independent complex variables. We will refer to them as holomorphic and anti-holomorphic respectively. We denote a scattering amplitude involving $n$ massless particles (all considered outgoing) with momenta $p_1 = \lam_1 \lamt_1,\, \dots ,\, p_n = \lam_n \lamt_n$ and helicities $s_1, \dots, s_n$ by $\mA_n \left( \left\lbrace \lam_1, \lamt_1\right\rbrace^{s_1}, \dots, \left\lbrace \lam_n, \lamt_n\right\rbrace^{s_n} \right)$. Consider an all-line holomorphic shift
\begin{align}
    \label{eq:CSWshift}
    \hat{\lam}_i = \lam_i + \alpha \, w_i X \qquad i = 1, \dots , n
\end{align}
where $X$ is an arbitrary reference spinor, $\alpha$ is the deformation parameter and the $w_i$ are chosen to satisfy momentum conservation
\begin{align}
  \sum_{i=1}^n  w_i \lamt_i = 0\,.
\end{align}
The deformed amplitude has the following large $\alpha$ behavior~\cite{Cohen:2010mi}
\begin{align}
\label{eq:CSWconstructibility1}
    \hat{\mA}_n \left(\alpha\right) \to \alpha^a \qquad \text{as } \qquad \alpha \to \infty \qquad \text{with }\qquad  2a = 4-n-c-\sum_{i=1}^n s_i\,,
\end{align}
where $c$ is the mass dimension of the product of couplings in the amplitude. The undeformed amplitude can be related to its residues at non-zero values of $\alpha$ via the residue theorem
\begin{align}
    \label{eq:CSW}
\mA_n \left(\alpha = 0\right) = \underset{\alpha = 0}{\oint} \frac{\dd \alpha}{\alpha} \, \mA_n \left(\alpha\right)  = -\sum_{j}   \underset{\alpha = \alpha_j}{\text{Res}} \left[ \frac{1}{\alpha}\mA_n \right].
\end{align}
In order to use the all-line shift recursion to compute amplitudes recursively, it is crucial that there be no contribution to~(\ref{eq:CSW}) from a residue at infinity.  From~(\ref{eq:CSWconstructibility1}) we see that this will be the case as long as
\begin{align}
\label{eq:CSWconstructability}
    4 - n - c - \sum_{i=1}^n s_i < 0\,.
\end{align}
When this holds, then the tree-level deformed amplitude only has simple poles which occur when the sum of a subset of the external momenta goes on shell. The amplitude factorizes into lower-point subamplitudes on the residue of the pole.

In the paper, our main interest lies in the collinear channel where we have
\begin{equation}\label{eq:diag123}
    \begin{tikzpicture}[baseline={([yshift=-1ex]current bounding box.center)}]
        \draw[thick] (0,0) -- (120:1) node[left]{$2^{s_2}$};
        \draw[thick] (0,0) -- (240:1) node[left]{$1^{s_1}$};
        \draw[thick] (0,0) node[above right]{$\ P_{I}^{-s_I}$} -- (0:3) node[above left]{$-P_{I}^{s_I}\ \,$} -- ++ (0,1) node[right]{$3^{s_3}$} (0:3) -- ++ (0.7,-0.7) node[right]{$\cdots$} -- ++ (0.3,-0.3);
        \filldraw[thick,fill=white] (0,0) circle (.3);
        \filldraw[thick,fill=white] (3,0) circle (.3);
    \end{tikzpicture}
\end{equation}
The value of $\alpha$ that corresponds to this channel is the one that makes the intermediate momentum $P_{I}$ go on shell:
\begin{align}
    \label{eq:collinearCSW}
    &\hat{P}_{I}^2 = \an{\hat{1}\hat{2}}\left[12\right] = 0 \qquad \implies\qquad \alpha = -\frac{\an{12}}{w_1 \an{X2}-w_2\an{X1}}\,,
\end{align}
and we can write
\begin{align}
\label{eq:one}
    &\lamh_1 = \frac{\an{X1}}{w_1 \an{X2}-w_2\an{X1}}\,\left(w_1\lam_2-w_2\lam_1\right)\,,  \qquad\lamh_2 =  \frac{ \an{X2}}{w_1 \an{X2}-w_2\an{X1}}\,\left(w_1\lam_2-w_2\lam_1\right)\,, \\
    &    \lamh_I = w_1 \lam_2 - w_2 \lam_1\,,  \qquad\qquad\qquad\quad\quad\quad\quad\quad\,\,\,\,\hat{\lamt}_I = \frac{\an{X1}\lamt_1+\an{X2}\lamt_2}{w_1\an{X2}-w_2\an{X1}}\,.
\label{eq:two}
\end{align}
We therefore have
\begin{align}
\label{eq:CSWrecursion}
&{\cal A}_n \left(\left\lbrace \lam_1, \lamt_1\right\rbrace^{s_1}, \left\lbrace \lam_2, \lamt_2\right\rbrace^{s_2}, \dots, \left\lbrace \lam_n, \lamt_n\right\rbrace^{s_n}\right) \\
\nonumber=& \sum_{s_I} \hat{\mA}_3  \left(\left\lbrace \lamh_1, \lamt_1\right\rbrace^{s_1}, \left\lbrace \lamh_2, \lamt_2\right\rbrace^{s_2},\left\lbrace \lamh_I, \hat{\lamt}_I\right\rbrace^{-s_I}\right)  \frac{1}{\an{12}\sq{12}} \hat{\mA}_{n-1} \left(\left\lbrace \lamh_I, \hat{\lamt}_I\right\rbrace^{s_I}, \dots, \left\lbrace \lamh_n, \lamt_n\right\rbrace^{s_n}\right)\\
&\nonumber \qquad\qquad + \left\lbrace\text{other channels}\right\rbrace
\end{align}
The other channels may include other collinear (for e.g. 34 collinear) as well as multiparticle poles.

The recursive computation of amplitudes is seeded by three-point amplitudes, which are completely fixed by Lorentz invariance and little group scaling to be\footnote{In~(\ref{eq:YMvertices}) and~(\ref{eq:newtwo}) we define certain $\kappa$'s with different overall normalizations compared to this standard.}
\begin{equation}
\label{eq:threepoints}
    A(1^{s_1},2^{s_2},3^{s_3}) = \left\{
    \begin{aligned}
        & \kappa_{s_1,s_2,s_3} \lbrack 12 \rbrack^{s_1+s_2-s_3} \lbrack 23 \rbrack^{s_2+s_3-s_1} \lbrack 31 \rbrack^{s_3+s_1-s_2} , && \text{if } s_1+s_2+s_3 > 0\,, \\
        & \kappa_{s_1,s_2,s_3} \langle 12 \rangle^{s_3-s_1-s_2} \langle 23 \rangle^{s_1-s_2-s_3} \langle 31 \rangle^{s_2-s_1-s_3} , && \text{if } s_1+s_2+s_3 < 0 \,.
    \end{aligned}
    \right.
\end{equation}
A crucial aspect of the holomorphic shift is that the holomorphic three-point amplitudes vanish. This follows from the proportionality of the three holomorphic spinors $\hat{\lambda}_1$, $\hat{\lambda}_2$ and $\hat{\lambda}_I$ in~(\ref{eq:one}), (\ref{eq:two}). Thus, only anti-holomorphic three-point amplitudes contribute in the collinear limit.

\bibliographystyle{JHEP}

\bibliography{main7}

\end{document}